\documentclass[11pt]{article}
\usepackage{amsmath,amssymb,color,graphics,epsfig,cite}
\usepackage{amsfonts}
\newcommand{\hoch}[1]{$\, ^{#1}$}

\newcommand{\be}{\begin{equation}}
\newcommand{\ee}{\end{equation}}
\newcommand{\bea}{\setlength\arraycolsep{2pt} \begin{eqnarray}}
\newcommand{\eea}{\end{eqnarray}}
\newcommand{\nn}{\nonumber}

\def\ft#1#2{{\textstyle{\frac{\scriptstyle #1}{\scriptstyle #2} } }}
\def\fft#1#2{{\frac{#1}{#2}}}

\def\0{{\sst{(0)}}}
\def\1{{\sst{(1)}}}
\def\2{{\sst{(2)}}}
\def\3{{\sst{(3)}}}
\def\4{{\sst{(4)}}}
\def\5{{\sst{(5)}}}
\def\6{{\sst{(6)}}}
\def\7{{\sst{(7)}}}
\def\8{{\sst{(8)}}}
\def\sst#1{{\scriptscriptstyle #1}}

\begin{document}

\begin{center}
{\Large {\bf Quadratic Curvature Correction and Its Breakdown\\ to Thermodynamics of Rotating Black Holes}}

\vspace{20pt}

Peng-Yu Wu\hoch{1} and H. L\"{u}\hoch{1,2}

\vspace{10pt}

{\it \hoch{1}Center for Joint Quantum Studies and Department of Physics,\\
School of Science, Tianjin University, Tianjin 300350, China }

\bigskip

{\it \hoch{2}Joint School of National University of Singapore and Tianjin University,\\
International Campus of Tianjin University, Binhai New City, Fuzhou 350207, China}

\vspace{40pt}

\underline{ABSTRACT}
\end{center}

We adopt the Reall-Santos method to obtain the quadratic curvature corrections to the thermodynamics of Myers-Perry rotating black holes in diverse dimensions. We consider the corrections in canonical and grand canonical ensembles, and also for fixed mass and angular momenta. We find that there exist inevitable divergences for generic rotation parameters so that the perturbative approach breaks down. We present a simple and universal formula to determine where these divergences could arise, and argue, using an explicit example, that these divergences are artefact of perturbation and may not exist in the full nonlinear theory.

\vfill{wupy2023@tju.edu.cn\ \ \  mrhonglu@gmail.com}
%\vfill {\footnotesize mrhonglu@gmail.com}

%{\footnotesize \hoch{*}Corresponding author}

\thispagestyle{empty}
\pagebreak

\tableofcontents
\addtocontents{toc}{\protect\setcounter{tocdepth}{2}}

\newpage

\section{Introduction}

Kerr metric \cite{Kerr:1963ud} is arguably the most important vacuum solution in General Relativity, with both theoretical and astrophysical importance. During the heat of string theory in the 80's, Myers and Perry generalized the Kerr metric to arbitrary dimensions \cite{Myers:1986un}. These solutions were further generalized to include a cosmological constant \cite{Gibbons:2004uw,Gibbons:2004js}, extending the earlier results of Carter \cite{Carter:1968ks} and of Hawking, Hunter and Taylor \cite{Hawking:1998kw}. The fact that the highly nonlinear Einstein's field equation admits these complicated but elegant solutions is quite a wonder, but the miracle stops when high-derivative gravities are involved.

Keeping the general covariance, higher-derivative extensions in terms of curvature tensors or their covariant derivatives are natural ways to modify Einstein gravity. The simplest example is the quadratic extension, involving three structures in general dimensions
\be
\Delta {\cal L} = \sqrt{-g} \big( \alpha R^{\mu\nu\rho\sigma} R_{\mu\nu\rho\sigma} +
\beta R^{\mu\nu} R_{\mu\nu} + \gamma R^2\big)\,.
\ee
These higher-derivative terms introduce massive scalar and spin-2 modes, and the latter is ghostlike. The theory was demonstrated to be renormalizable \cite{Stelle:1976gc,Stelle:1977ry}, if the ghosty spin-2 mode is present. The usual black hole no-hair theorem holds for the massive scalar mode \cite{Nelson:2010ig,Lu:2015cqa}, but black holes with massive spin-2 hair exist and they were numerically constructed \cite{Lu:2015cqa,Lu:2015psa}, where the first law involving the Wald entropy \cite{Wald:1993nt} was demonstrated.

The alternative approach is to treat the higher-derivative gravities as the low-energy effective theories of some ultraviolet complete quantum gravity. In this perturbative approach, the quadratic extension is the leading-order correction. A massive mode, whose mass is inversely proportional to the coupling constant, is necessarily decoupled from the spectrum. One can further perform field redefinitions on the metric order by order, namely
\be
g_{\mu\nu} \rightarrow g_{\mu\nu} + c_1 R_{\mu\nu} + c_2 R g_{\mu\nu} + \cdots\,,
\ee
to set the coupling constants $\beta$ and $\gamma$ to zero, or to form any other combinations, such as the Gauss-Bonnet or Weyl-squared terms, while the $\alpha$-term is invariant. Since we consider the higher-order corrections to the Myers-Perry metrics, which are Ricci flat, the $R^2$ and $R^{\mu\nu} R_{\mu\nu}$ terms give no contribution to the equation of motion. Thus the above field redefinition is not necessary. Only the $\alpha$ term will contribute and furthermore, this $\alpha$ could be the coupling of either the Gauss-Bonnet or Weyl-squared term.

Unfortunately, even in this perturbative approach, exact solutions of rotating black holes are hard to come by, since the metric depends not only on the radial coordinate, but also on a plethora of latitudinal coordinates. This is because in general $D$ dimensions, there can be $[(D-1)/2]$ independent rotations along the longitudinal Killing directions. It is worth mentioning that in odd dimensions, the metric becomes cohomogeneity one, depending only on the radial coordinate, when all the angular momenta are equal. Such rotating black holes under the quadratic curvature perturbation at the $\alpha$ order were constructed in $D=5$ \cite{Ma:2020xwi} and $D=7$ \cite{Mao:2023qxq}, and the latter showed that even the outer horizon could be destroyed by the perturbation.

Recently, Reall and Santos (RS) developed a technique of evaluating the leading-order correction to black hole thermodynamics without having to solve the corresponding solution \cite{Reall:2019sah}. The technique is based on the quantum statistic relation \cite{Gibbons:1978ac}. They showed that the corrected Euclidean action is simply given by higher-derivative bulk action, evaluated on the leading-order solution. Not only is there no need to calculate the corrections to the leading-order solution, but also it is unnecessary to deal with the Gibbons-Hawking surface terms associated with the higher-derivative bulk terms. This technique has been heavily employed in the subject of the weak gravity conjecture\cite{Arkani-Hamed:2006emk,Kats:2006xp,Cheung:2018cwt,Hamada:2018dde,
Cheung:2019cwi,Arkani-Hamed:2021ajd,Aalsma:2022knj,Horowitz:2024dch} in the context of effective field theory, and also in supergravities \cite{Ma:2021opb,Ma:2022gtm} and gauged supergravities \cite{Bobev:2022bjm,Cassani:2022lrk,Ma:2024ynp}. The RS method belongs to a series of sequence, where, the $(n+1)$'th order correction requires only solutions up to and including the $n$'th order \cite{Ma:2023qqj}. In this regard, the starting sequence is the usual first law of black hole thermodynamics, which can be derived, based on the Wald formalism \cite{Wald:1993nt}, without needing even to know the full leading-order solution. When a cosmological constant are involved, the situation is much more subtle, the original RS method indeed gets rid of the need for finding the corrected solution, it still requires the knowledge of the surface terms, which can be intensively involved for higher-derivative gravities. Improved versions have been developed to remove the need of these surface terms \cite{Xiao:2023two,Hu:2023gru}.

In this paper, we study the quadratic curvature corrections to the thermodynamics of Myers-Perry black holes in general dimensions. The motivation is two fold. One is that the results are elegant and worth presenting. The second is the observation that the perturbative approach breaks down at certain points in the full parameter space. By breakdown, we mean that the corrected thermodynamic quantities become divergent at certain temperature. Such breakdown was know in zero temperature limit of a system with fixed mass and charges \cite{Cheung:2018cwt}, but we find that it can arise at nonzero temperature as well. In this paper, we shall understand how this divergence arises and give a universal simple formula to determine where this divergence could emerge.

The paper is organized as follows. In section \ref{sec:mp}, we review the Myers-Perry black holes and their thermodynamics. We also find that in odd dimensions, when one and only one of the angular momenta vanishes, the extremal limit leads to an interesting decoupling geometry, with an unusual AdS$_3$ embedding. In section \ref{sec:G}, we apply the RS method and obtain the thermodynamic corrections in the grand canonical ensemble, where the thermodynamic variables are temperature and angular velocities. Interestingly, we find that there is no divergence when all angular momentum are equal, but divergence emerges for generic parameters of angular momenta. We give reasons why these divergences emerge and determine where they do. In sections \ref{sec:F} and \ref{sec:MJ}, we study the thermodynamics in canonical ensemble, and also in fixed mass and angular momenta. We find that the divergence is a commonplace occurrence, typically at non-zero temperature. In section \ref{sec:D4}, we consider cubic curvature corrections in $D=4$, since the quadratic correction is trivial. In section \ref{sec:artefact}, we consider Einstein-Gauss-Bonnet gravity coupled to Maxwell field, where an exact solution of charged black hole exists. We use this example to illustrate that the divergence is an artefact of the perturbative approach. There is no such divergence at the full nonlinear level. We conclude the paper in section \ref{sec:con}. In appendix, we give corrections to all the thermodynamic quantities in the grand canonical ensemble.

\section{Thermodynamics of the Myers-Perry black holes}
\label{sec:mp}

\subsection{Thermodynamic variables}

Myers-Perry rotating black holes are described by Ricci-flat metrics that are asymptotic to Minkowski spacetimes in general $D$ dimensions. They are higher-dimensional generalizations of the four-dimensional Kerr metric. The asymptotically-Minkowski metric was later generalizaed to (A)dS rotating black holes, which are themselves generalization of four-dimensional Carter and five-dimensional Hawking-Hunter-Taylor (A)dS rotating black holes. In this paper, we adopt the convention of \cite{Gibbons:2004ai}, where the black hole thermodynamic quantities of Kerr-AdS black holes in general dimensions were obtained. We therefore shall not give explicit metrics. To obtain the thermodynamics of Myers-Perry black holes, we simply use those in \cite{Gibbons:2004ai}, but set the cosmological constant to zero. In $D$ dimensions, there can be at most $N=[(D-1)/2]$ independent orthogonal rotations; therefore, the metrics contain $(N+1)$ parameters, namely $(m, a_1, a_2,\ldots, a_N)$, parameterizing the mass and $N$ independent angular momenta. They are
\be
M_0=\frac{(D-2)\Sigma_{D-2}}{8 \pi} m\,,\qquad J^i_0=\frac{\Sigma_{D-2}}{4\pi} m a_i\,,\qquad i=1,2,\ldots, N\,,\label{M0J0}
\ee
where $\Sigma_D = \fft{2\,\pi^{\fft12(D+1)}}{\Gamma(\fft12(D+1)}$ is the volume of the round $D$-sphere. Note that we use subscript ``0'' to denote thermodynamic quantities in solutions of Einstein gravity. For sufficiently large mass $M_0$, there can exist an event horizon $r_0$. The $N+1$ parameters and $r_0$ are related by the horizon condition
\be
m=\ft12 r_0^{\epsilon-2} \prod_{i} (r_0^2 + a_i^2)\,,\qquad \epsilon=D-2N-1\,.\label{m}
\ee
It is advantageous to introduce $\epsilon$, which is 0 or 1 for $D$ = odd or even respectively. It is worth commenting here that for the rotating black hole with $(N+1)$ parameters $(m, a_i)$, the above equation also give all possible roots associated with Killing horizons. However, there can be either just one positive $r_+$, or at most two positive roots, $r_-\le r_+$. The position $r_+$ is the event horizon and $r_-$ gives the inner Cauchy horizon.

The remaining black hole thermodynamic quantities are the temperature, entropy and the angular velocities conjugate to the angular momenta. They are
\be
T_0=\fft{1}{2\pi}\Big( \sum_{i}\frac{r_0}{r_0^2+a_i^2}\,-\frac{1}{(\epsilon+1)r_0}\Big),\qquad
S_0=\fft{\Sigma_{D-2}}{4r_0^{1-\epsilon}} \prod_{i}(r_0^2+a_i^2)\,,\qquad \Omega_0^i =  \fft{a_i}{r_0^2 + a_i^2}\,.\label{tsomega}
\ee
The Gibbs free energy $G_0$ and the Euclidean action $I_0$ are
\be
I_0=\beta G_0\,,\qquad G_0\equiv M_0-T_0 S_0 -\sum_{i} \Omega^i_0 J^i_0=\fft{\Sigma_{D-2}\, m}{8\pi}\,.
\ee
The Gibbs free energy is the thermodynamic potential for grand canonical ensemble with thermodynamic variables $(T_0, \Omega_0^i)$. Compared to \eqref{M0J0}, we have
\be
M_0 = (D-2) G_0\,.\label{M0G0}
\ee
This relation is a consequence of the Smarr or Smarr-like relations associated with $M_0$ and $G_0$. Each thermodynamic quantity scales homogeneously under $r_0\rightarrow \lambda r_0$ and $a_i\rightarrow \lambda a_i$ with appropriate scaling dimensions
\be
[M_0]=[G_0]=D-3\,,\qquad [T_0]=[\Omega_0^i]=-1\,,\qquad [S_0]=[J_0^i] = D-2\,.\label{scalingd}
\ee
Thus from their respective ``first law'' of $M_0$ and $G_0$, we can deduce the two algebraic relations
\be
(D-3)M_0 =(D-2)( T_0 S_0 + \sum_{i} \Omega^i_0 J_0^i)\,,\qquad
(D-3)G_0 = T_0 S_0 + \sum_{i} \Omega^i_0 J_0^i \,,
\ee
which lead to \eqref{M0G0}. Finally, the Helmholtz free energy is
\be
F_0\equiv M_0 - T_0 S_0\,,
\ee
which is the thermodynamic potential for the canonical ensemble with thermodynamic variables $(T_0, J_0^i)$. Analogous Smarr-like relation can be obtained, but there is no special simple direct relation between the free energy and mass.

\subsection{Extremal limit and mass-angular momentum relation}

A particular theoretic interest is the zero temperature extremal limit, $T_0=0$. When all the angular momentum parameters $a_i$ are nonzero, such a limit exists for some positive $r_0$. In the extremal limit, the mass and angular momenta are no longer independent variables, but they are related by an algebraic relation in terms of some polynomial functions, which are in general too complicated to present. Here we give three lower-dimensional examples:
\bea
D=4:&& M_{\rm ext}=\sqrt{|J|}\,,\nn\\
D=5:&& M_{\rm ext}=\frac{3 \pi^{1/3}}{2}\left(\frac{|J_1|+|J_2|}{2}\right)^{2/3},\nn\\
D=6:&& M_{\rm ext}=\frac{2 \sqrt[8]{2} \sqrt[4]{\pi } \sqrt[8]{-J_1^6+33 J_2^2 J_1^4+33 J_2^4 J_1^2-J_2^6+\left(J_1^4+14 J_2^2 J_1^2+J_2^4\right){}^{3/2}}}{3^{5/8}}\,.\label{lowdmext}
\eea
When all angular momenta are equal, namely $J_i = \fft{J}{N}$, the general dimensional mass-angular momentum relation in the extremal limit becomes presentable:
\be
M_{\rm ext} = \xi J^{\frac{D-3}{D-2}}\,,\qquad
\xi=\fft{(D-2)\Big(\fft{\Sigma_{D-2}}{32\pi}\Big)^{\fft{1}{D-2}}2^{\frac{\epsilon }{2 (D-2)}} (\epsilon +1)^{\frac{D-1-\epsilon}{2(D-2)}}}{(D-\epsilon -1)^{\frac{D-5+\epsilon}{2 (D-2)}} (D (\epsilon +1)-\epsilon  (\epsilon +2)-3)^{\frac{D-3}{2 (D-2)}}}\,.\label{xi}
\ee
For odd dimensions, corresponding to $\epsilon=0$, the relation was given in \cite{Mao:2023qxq}. In the extremal limit, there exists decoupling limit that magnifies the horizon geometry which contains a $U(1)$ bundle of AdS$_2$ \cite{Bardeen:1999px}. The geometry plays an important role in the Kerr/CFT proposal \cite{Guica:2008mu}, which was subsequently generalized to diverse dimensions \cite{Lu:2008jk}.

\subsection{Singular extremal limit and an AdS$_3$ embedding}

Curiously, in the general $D=5,6$ examples of \eqref{lowdmext}, if we set $J_2=0$, the mass does not vanish for $D=5$, but vanishes for $D=6$. Indeed, for even dimensions, as long as one or more than one angular momentum vanishes, there can be no extremal limit. For odd dimensions, extremal limit is also possible if only one angular momentum vanishes. Without loss of generality, we set $J_N=0$, (i.e.~$a_N=0$,) for which case, we have
\be
T_0 = \fft{r_0}{2\pi} \sum_{i=1}^{N-1} \fft{1}{r_0^2 + a_i^2}\,.
\ee
Thus the temperature goes to zero when we have $r_0\rightarrow 0$. In this limit, the mass/angular momentum relation becomes
\be
M=\ft12(D-2) \Big(\fft{\Sigma_{D-2}}{8\pi}\Big)^{\fft{1}{D-2}} \prod_{i=1}^{N-1} |J_i^0|^{\fft{2}{D-2}}\,,\qquad J_N=0\,.\label{nearlymj}
\ee
The extremal metric also allows a decoupling limit, which starts with the scaling
\be
r\rightarrow \lambda r\,,\qquad \phi_i \rightarrow \phi_i + \fft{t}{\lambda a_i}\,,\qquad
t\rightarrow \fft{t}{\lambda}\,,\qquad \phi_N \rightarrow \fft{\phi_N}{\lambda}\,,
\ee
followed by sending $\lambda\rightarrow 0$. The resulting $D=2N+1$ metric, with $a_N=0$, becomes
\bea
ds^2 &=& \mu_N^2 ds_{\rm AdS_3}^2 + \sum_{i=1}^{N-1} a_i^2\bigg( d\mu_i^2 + \mu_i^2\Big(1+ \fft{\mu_i^2}{\mu_N^2}\Big) d\phi_i^2\bigg) + \sum_{i<j}^{N-1} 2a_i a_j \fft{\mu_i^2\mu_j^2}{\mu_N^2} d\phi_id\phi_j\,,\nn\\
ds_{\rm AdS_3}^2 &=& \fft{\ell^2 dr^2}{r^2} - \fft{r^2}{\ell^2} dt^2 + r^2 d\phi_N^2\,,\qquad \fft{1}{\ell^2} = \sum_{i=1}^{N-1} \fft{1}{a_i^2}\,,\qquad
\sum_{i=1}^N \mu_i^2=1\,.\label{ads3}
\eea
The metric has only the curvature singularity at the ``equator'' $\mu_N=0$. In particular, the conformally scaled metric, namely $ds^2/\mu_N^4$, is absent from any curvature singularity. This unusual AdS$_3$ embedding is very different from the near-horizon extreme Kerr studied in \cite{Bardeen:1999px,Lu:2008jk}.

\subsection{Thermodynamic instability}

The thermodynamic instability of the Myers-Perry black  holes was studied in \cite{Monteiro:2009tc} and some low-dimensional examples were explicitly demonstrated. The method is based on studying the positivity of some appropriate thermodynamic metric. We find that it is worth reviewing this method since there is a simple and analytic way to determine the thermodynamic metric \cite{Liu:2010sz}. For a given ensemble with thermodynamic potential $\Phi$ of variables $Q^i$, satisfying the first law
\be
d\Phi = \sum_i \mu^i dQ^i\,,
\ee
the metric based on the Hessian matrix of $\Phi$ is simply given by \cite{Liu:2010sz}
\be
ds^2_\Phi \equiv \sum_{i,j}\fft{\partial^2 \Phi}{\partial Q^i\partial Q_j} dQ^i dQ^i =\sum_{i}
d\mu^i\, dQ^i\,.
\ee
This allows one to use horizon radius, charge and angular momentum parameters to write analytically the thermodynamic metric instead of $Q^i$, for which, $\Phi$ generally does not have a close analytic expression. The whole point of introducing thermodynamic metric is precisely to study the ``coordinate'' independent properties.

In thermodynamics, two particular metrics are important. One is the Weinhold metric based on the Hessian matrix of mass $M_0$ \cite{Weinhold:1975xej}, and the other is the Ruppeiner metric based on the Hessian matrix of entropy $S_0$ \cite{Ruppeiner:1979bcp}. For the general rotating black holes, they are
\bea
ds_{M_0}^2 &=& dT_0 dS_0 + \sum_i d\Omega_0^i dJ_0^i\,,\nn\\
ds_{S_0}^2 &=& d\left( \fft{1}{T_0}\right)  dM_0 - \sum_i d\left(\fft{\Omega_0^i}{T_0}\right) dJ_0^i=
-\fft{1}{T_0} ds_{M_0}^2\,.
\eea
The thermodynamic potential associated with variables mass and angular momenta is the entropy, and its thermodynamic metric is conformal to that of the mass. The thermodynamic stability thus requires that $ds_{M_0}^2$ must be nonnegative, or equivalent $ds_{S_0}^2$ is non-positive. For the Myers-Perry black holes in general dimensions, the thermodynamic quantities can all be expressed in terms of horizon $r_0$ and rotation parameters $a_i$'s. Thus we can write the metric formally as
\be
ds_{M_0}^2 = g_{\mu \nu} dx^\mu dx^\nu\,,\qquad x^\mu = (r_0, a_i)\,.\label{dsm0}
\ee
For thermodynamic stability, the metric must be nonnegative for all allowed values of $x^\mu$. For the rotating black holes in even dimensions, we can easily show that this is not possible, since the trace of the metric has a simple expression
\be
{\rm tr} (g_{\mu\nu}) = -\fft{4r_0}{D-2}\, \fft{\partial^2 (r_0 M_0)}{\partial(r_0^2)\partial(r_0^2)}\,,\qquad (\hbox{for even}\,\, D,)\label{traceevend}
\ee
which is manifestly negative. (In $D=4$, it is simply zero, since $r_0M_0$ is proportional to $r_0^2$.) We thus establish that the rotating black holes in even dimensions are all thermodynamically unstable.

The situations in odd dimensions are more subtle, where the relation \eqref{traceevend} does not apply. Furthermore, the quantity ${\rm tr}(g_{\mu\nu})$ can be both positive and negative. In odd dimensions, we can define $k\times k$ matrix
\be
{\cal M}^k_{\mu\nu} = g_{\mu\nu},\qquad {\mu,\nu}=0,1,\ldots, k-1\,.
\ee
The thermodynamics stability requires nonnegativity of the metric \eqref{dsm0}, which in turn implies that $\det({\cal M}^k)\ge 0$ for all $k$. This is a very strong condition and in five dimensions, we have
\be
{\rm det}({\cal M}^{k=3}) = -\frac{3 \pi ^4 \left(a_1^2+r_0^2\right)^2 \left(a_2^2+r_0^2\right)^2}{32 r_0^7} T_0\,.
\ee
In other words it is negative for $T_0>0$. When $T_0=0$, we find that ${\rm det} ({\cal M}^{k=2})=-3 \pi ^2/4$. We thus conclude that the five-dimensional rotating black hole is also thermodynamically unstable. For even higher odd dimensions, we have to use numerical analysis, and we find that for any allowed parameters (such that $T_0\ge 0$,) there is at least one  $\det({\cal M}^k)$ that is negative, indicating that Myers-Perry black holes in odd dimensions are also all thermodynamically unstable. Specifically, we have checked both $D=7$ and $D=9$. In each case, we randomly select a real number from 0 to $10^6$ for each item in the $(r_0,a_i)$ set. We find that $T_0$ and all ${\cal M}^k$'s can each be either positive and negative, but none of the randomly selected $(r_0,a_i)$ values would give all positive ${\cal M}^k$'s simultaneously. We have explicitly checked billions of the $(r_0,a_i)$ datasets using a computer program and found no exception, strongly indicating that the rotating black holes in odd dimensions are thermodynamically unstable.

As we shall demonstrate in section \ref{sec:MJ}, the specific heat capacity can be positive for low temperature including the extremal limit. Therefore, the thermodynamic can be stable if we fix all the angular momenta to some specific values so that they do not involve in the thermodynamic process.

\section{Corrections in the grand canonical ensemble}
\label{sec:G}

We now study the leading-order corrections to the thermodynamic quantities of the Myers-Perry black holes discussed in the previous section by the quadratic curvature invariants. A traditional approach is to perturb the background metric $\bar g_{\mu\nu}$ to $g_{\mu\nu} = \bar g_{\mu\nu} + \delta g_{\mu\nu}$. As we have discussed in the introduction, only the $\alpha$ term will contribute, so $\delta g_{\mu\nu}$ is of order $\alpha$. This approach can be effective for spherically-symmetric and static backgrounds, where the metrics are of cohomogeneity one. For general $D$-dimensional Myers-Perry metrics, this is an almost impossible task, except when all the angular momenta are equal in odd spacetime dimensions, where the metrics become cohomogeneity one again. Indeed explicitly constructions of such metrics in five \cite{Ma:2020xwi} and seven dimensions \cite{Mao:2023qxq} were carried out.

However, it is not necessary to know the complete solution in order to derive the first law of black hole thermodynamics. The Wald formalism indicates that the first law that connects the data in both asymptotic infinity and horizon regions can be derived by the a few leading-order power series expansions in both regions. As was discussed in the introduction, it was recently established that the leading-order correction to the black hole thermodynamics can be obtained by having only to know the original uncorrected solution \cite{Reall:2019sah}. These are in fact two low-lying examples in the general series that the $(n+1)$'th correction to the thermodynamics requires only solutions up to and including the $n$'th order \cite{Ma:2023qqj}.

\subsection{Explicit corrections}

The procedure is based on the quantum statistic relation, namely the Euclidean action of Einstein gravity is proportional to the Gibbs' free energy of grand canonical ensemble, where temperature and the angular velocities conjugate to the angular momenta are thermodynamic variables \cite{Gibbons:1978ac}. For the rotating black holes, we have
\be
I = \beta G(T, \Omega_i)\,,\qquad \beta = \fft{1}{T}\,.
\ee
Applying the RS method to our case, the total Euclidean action is given by
\be
I_{\rm tot} = I_0 + \Delta I\,,\qquad \Delta I = \fft{\alpha }{16\pi} \int_{\cal M} \sqrt{g}\, R^{\mu\nu\rho\sigma} R_{\mu\nu\rho\sigma}\Big|_{\bar g_{\mu\nu}}\,.\label{eucact}
\ee
The integration is over the whole space on and outside of the horizon. The Gibbs free energy is simply $G=G_0 + \Delta G$, with $\Delta G= T_0\, \Delta I$. Thus we have
\be
\Delta G=\left\{
    	\begin{aligned}
    	&(-1)^{N}\frac{(D-3)\Sigma_{D-2}}{4\pi} \alpha\, m^2  \sum_{i}^{N}\frac{r_0^4-6a_i^2r_0^2+a_i^4}{b_i(r_0^2+a_i^2)^3}\,,\qquad &D=\hbox{odd;}\\
    	& (-1)^{N}\frac{(D-3)\Sigma_{D-2}}{\pi} \alpha\, m^2  \sum_{i}^{N}\frac{r_0\left(r_0^2-a_i^2 \right) }{ b_i(r_0^2+a_i^2)^3}\,,\qquad &D=\hbox{even,}
    	\end{aligned}
    \right.\label{DeltaG}
\ee
where $b_i=\prod\limits_{j\neq i}(a_i^2-a_j^2)$. Note that the evaluation of the volume integration in \eqref{eucact} can be very involved as the dimensions increase. We first conjectured \eqref{DeltaG} from the explicit evaluation of \eqref{eucact} for $D=5,6,7$ dimensions, and then verified the formulae for $D=8,9,10,11$ dimensions.

Since in the grand canonical ensemble, $(T, \Omega^i)$ are thermodynamic variables, it is natural to treat them as fixed under the perturbation, thus we have
\be
T=T_0\,,\qquad \Omega^i= \Omega^i_0\,,\qquad M=M_0+ \Delta M\,,\qquad
J^i=J_0^i + \Delta J^i\,.
\ee
The perturbed Gibbs free energy satisfies its own first law
\be
d\Delta G= -\Delta S\, d T_0 - \sum_i\Delta J^i\, d\Omega_0^i\,.\label{pertfl}
\ee
This allows us to read off the perturbed entropy $\Delta S$ and $\Delta J^i$. (See appendix \ref{app:thermo} for details.) The perturbed mass is then
\be
\Delta M = \Delta G + T_0 \Delta S + \sum_i\Omega_0^i \Delta J^i\,,\qquad d\Delta M = T_0 d\Delta S + \sum_i\Omega_0^i d\Delta J^i\,.
\ee
However, we can use the Smarr-like relations to determine the $\Delta M$ more straightforwardly.  Owing to the fact that $[\alpha]=2$, but it does not participate in the thermodynamic variation, it follows $\Delta G$ has the scaling dimension of $(D-5)$ under constant scaling of $r$ and $a_i$ discussed above \eqref{scalingd}. Similarly, the scaling dimensions of $(\Delta S, \Delta J^i)$ are both $(D-4)$. We thus have the Smarr-like relations
\be
(D-5) \Delta G = T_0 \Delta S + \sum_i\Omega_0^i \Delta J^i\,,\qquad
(D-5) \Delta M= (D-4)\Big(T_0 \Delta S + \sum_i\Omega_0^i \Delta J^i)\,.
\ee
These lead straightforwardly to
\be
\Delta M = (D-4) \Delta G.\label{granddeltam}
\ee
This implies that in the grand canonical ensemble, both perturbed $\Delta G$ and $\Delta M$, following from \eqref{DeltaG}, are regular in the whole parameter space. For these two quantities, the perturbation theory is well defined.

\subsection{Extremal limit}

\subsubsection{All equal and nearly all equal angular momenta}

The general solution has $(N+1)$ independent parameters $(r_0, a_i)$. In the extremal $T_0=0$ limit, $r_0$ is fixed as a function of $a_i$. It follows that $M$ and angular momenta are no longer independent, but a function of $J^i$. The general case can be too complicated to present. We shall first consider two special cases. The first is when the angular momenta are all equal, i.e.~$a_i=a$. The uncorrected
thermodynamic variables are
\bea
M_0 &=& \fft{(D-2)\Sigma_{D-2}}{16\pi} r_0^{\epsilon-2} (r_0^2 + a^2)^N\,,\qquad
J_0=N J_0^i=\fft{2N}{D-2}M_0 a\,,\qquad \Omega_0=\fft{a}{r_0^2 + a^2}\,,\nn\\
T_0 &=& \fft{1}{2\pi} \Big(\fft{N r_0}{r_0^2 + a^2} - \fft{1}{(\epsilon+1) r_0}\Big)\,,\qquad S_0 = \fft{\Sigma_{D-2}}{4r_0^{1-\epsilon}} (r_0^2 + a^2)^N\,.
\eea
For odd dimensions $D=2N+1$, the leading $\alpha$ perturbation with fixed $(T_0, \Omega_0)$ are given by
\bea\label{J1}
\Delta M &=& (D-4) \Delta G=-\alpha\fft{(D-4)(D-3)\Sigma_{D-2}}{16\pi r_0^4 \left(a^2+r_0^2\right)^{\frac{5-D}{2}}}
\left(a^4-2(2D-3) a^2 r_0^2+(D-2)^2 r_0^4\right),\nn\\
\Delta J &=& -\frac{\alpha  N  (D-1)(D-3) a\left(a^2+r_0^2\right)^{\frac{D-5}{2}}\Sigma_{D-2}}{ 16 \pi  r_0^4 \left(2 a^2+(D-3) r_0^2\right)}\Big(
2 a^6-7 a^4 (D-3) r_0^2\cr
&&-2\left(D^2-5 D+1\right) a^2 r_0^4 +(D-4) \left(D^2-3 D+4\right) r_0^6\Big),\nn\\
\Delta S &=& -\frac{\alpha  (D-3) \left(a^2+r_0^2\right){}^{\frac{D-5}{2}}\Sigma_{D-2}}{4 r_0^3 \left(2 a^2+(D-3) r_0^2\right)}\Big(4 a^6-a^4 (7 D-11) r_0^2\cr
&&-2 (D-3) (2 D-3) a^2 r_0^4 +(D-5) (D-2)^2 r_0^6\Big).
\eea
In even $D=2 N+2$ dimensions, we have
\bea\label{J2}
\Delta M &=& -\frac{\alpha  (D-4) (D-3) (D-2) \Sigma_{D-2} \left((D-2) r_0^2-2 a^2\right)}{16 \pi  r_0 \left(a^2+r_0^2\right)^{\frac{6-D}{2}}}\,,\nn\\
\Delta J &=& \frac{\alpha  (D-4) (D-3) (D-2)\Sigma_{D-2}\, a  \left(2 a^4+a^2 (D-2) r_0^2-\left(D^2-3 D+4\right) r_0^4\right)}{ 16 \pi  r_0 \left(a^2+(D-3) r_0^2\right)\left(a^2+r_0^2\right)^{\frac{6-D}{2}}}\,,\nn\\
\Delta S &=&-\frac{\alpha  (D-3) (D-2) \Sigma_{D-2}\left(-2 a^4+a^2 (8-3 D) r_0^2+\left(D^2-7 D+10\right) r_0^4\right)}{4 \left(a^2+(D-3) r_0^2\right)
\left(a^2+r_0^2\right)^{\frac{6-D}{2}} }\,.
\eea
These quantities are well defined for all $r_0>0$, implying that there is no breakdown in the perturbative approach for these generic thermodynamic variables.
In the extremal limit, with $T_0=0$, we have
\be
M_{\rm ext} = \xi J^{\fft{D-3}{D-2}} (1 + \eta \alpha J^{-\fft{2}{D-2}})\,,\label{allequalmj}
\ee
where $\xi$ is given in \eqref{xi} and we find that $\eta$ is
\be
\eta = \left\{
  \begin{array}{ll}
    \fft{3D-11}{2(D-2)} (D-1)^{\fft{3}{D-2}} (D-3)^{\fft{D-1}{D-2}}\left(\frac{\Sigma _{D-2}}{32 \pi }\right)^{\frac{2}{D-2}}, &\qquad D = \hbox{odd;} \\
    4^{\frac{1}{D-2}} (D-4) (D-3)^{\frac{D-1}{D-2}} (D-2)^{\frac{4-D}{D-2}} \left(\frac{\Sigma_{D-2}}{32 \pi }\right)^{\frac{2}{D-2}} , &\qquad  D= \hbox{even.}
  \end{array}
\right.
\ee
The odd dimensional result was obtained in \cite{Mao:2023qxq}.

As we have discussed, in odd dimensions, there is an additional extremal limit when one and only one of the angular momentum vanishes. For simplicity, we set all the non-vanishing angular momenta equal, namely
\be
a_N=0\,,\qquad a_i=a\,,\qquad i=1,2,\ldots\,,N-1\,.\label{nearly}
\ee
In this paper, we call this as ``{\it nearly} all equal angular momenta.'' In this case, we have
\bea
M_0 &=& \fft{(D-2)\Sigma_{D-2}}{16\pi} (r_0^2 + a^2)^{N-1}\,,\quad
J_0=(N-1) J_0^i=\fft{2(N-1)}{D-2}M_0 a\,,\quad \Omega_0=\fft{a}{r_0^2 + a^2}\,,\nn\\
T_0 &=& \fft{(N-1)r_0}{2\pi(r_0^2 + a_i^2)}\,,\qquad S_0 = \fft{\Sigma_{D-2}}{4}r_0 (r_0^2 + a^2)^{N-1}\,.
\eea
The quadratic curvature corrected thermodynamic variables are
\bea\label{J3}
\Delta M &=& -\frac{\alpha  (D-4) (D-3)\Sigma_{D-2}}{16 \pi  r_0^2} \left(a^2+r_0^2\right){}^{\frac{D-7}{2}} \left(a^4+2 a^2 r_0^2+(D-2)^2 r_0^4\right)\,,\cr
\Delta J &=& -\frac{a \alpha (N-1)\Sigma_{D-2}}{8 \pi  r_0^2 \left(a^2+r_0^2\right)^{\frac{7-D}{2}}} \left((D-7)a^2(a^2 +2 r_0^2)+(D-4) \left(D^2-3 D+4\right) r_0^4\right),\nn\\
\Delta S &=&-\frac{\alpha  \Sigma_{D-2}}{4 r_0^3 \left(a^2+r_0^2\right)^{\frac{7-D}{2}}} \Big(2 a^6+a^4 (D-1) r_0^2-2 a^2 \left(D^2-5 D+7\right) r_0^4\cr
&&+(D-5) (D-2)^2 r_0^6\Big).
\eea
In the extremal limit $T_0\rightarrow 0$, ($r_0\rightarrow 0$), all the perturbed thermodynamic quantities ($\Delta M$, $\Delta J$, $\Delta S$) are divergent. The leading mass/angular momenta relation \eqref{nearlymj} therefore cannot be extended to the next order, in contrast to \eqref{allequalmj} for the case of all equal angular momenta. This breakdown of the perturbation theory may be too special since the near-horizon geometry, given in \eqref{ads3} with $(N-1)$ $a_i$'s all equal, is actually singular. After all, when all angular momenta are equal, we have seen no divergence at all! We thus examine further examples below.
\subsubsection{$D=5$}

Another example we would like to focus in this grand canonical ensemble is the five dimensional rotating black hole. We have $\Delta M=\Delta G$ and the corrections of the remainder thermodynamic quantities are
\bea
\Delta J^1 &=&  -\fft{\alpha a_1}{4 T_0\, r_0^ 5\prod_{i} (r_0^2 + a_i^2)^2}\Big(
7 r_0^{12}+\left(8 a_2^2-2 a_1^2\right) r_0^{10}-\left(a_1^4+25 a_1^2 a_2^2-11 a_2^4\right) r_0^8\cr
&&-2 a_2^2 \left(5 a_1^4+2 a_1^2 a_2^2-a_2^4\right) r_0^6-a_1^2 a_2^2 \left(a_1^4-15 a_1^2 a_2^2+5 a_2^4\right) r_0^4 + 6 a_2^4 a_1^6 r_0^2+a_1^6 a_2^6\Big)\,,\cr
\Delta J^2 &=& \Delta J^1 \Big|_{a_1\leftrightarrow a_2}\,,\cr
%%%%%
\Delta S &=& \fft{\pi \alpha}{2 T_0\, r_0^4\prod_{i} (r_0^2 + a_i^2)^2}\Big(
7 \left(a_1^2+a_2^2\right) r_0^{10}-2 \left(a_1^4-15 a_2^2 a_1^2+a_2^4\right) r_0^8\cr
&&-\left(a_1^2+a_2^2\right) \left(a_1^4+a_2^4\right) r_0^6-28 a_1^4 a_2^4 r_0^4-5 a_1^4 a_2^4 \left(a_1^2+a_2^2\right) r_0^2+2 a_1^6 a_2^6
\Big)\,.
\eea
We thus see that all these thermodynamic corrections are divergent in the extremal limit with $T_0\rightarrow 0$. This implies the breakdown of the perturbative approach at zero temperature, where we cannot even have perturbative mass/charge relation for the extremal five-dimensional rotating black hole with generic angular momenta. This is a counterexample to \cite{Horowitz:2024dch} that suggested that there would have no breakdown for fixed zero temperature. Interestingly, when the two angular momenta are equal, corresponding to $a_1=a_2$, the numerators of the above quantities all have a factor that cancels the denominator $T_0$. We therefore have no divergence for all equal angular momenta, and this five dimensional case is just one of the general cases discussed earlier.

Both divergent examples considered above seem to indicate problems at zero temperature. When more than two angular momenta in odd dimensions, or one angular momentum in even dimensions vanishes, there is no extremal limit. Are these black holes immune from the perturbation breakdown? We examine further examples below.

\subsection{Perturbative breakdown of at non-zero critical temperature}

In the previous subsection, the perturbative approach breaks down at zero temperature for the five-dimensional rotating black hole, except when two angular momenta are equal. Such breakdowns in fact are a common occurrence in higher dimensions as well, although the critical temperature happens not be zero. We shall study this phenomenon in this subsection and determine the critical temperature. For simplicity, we first consider a simpler case with all angular momenta but one vanish. The non-vanishing thermodynamic quantities of the Ricci-flat rotating black hole are
\bea
&&M_0=\frac{(D-2)\Sigma_{D-2}}{16 \pi} r_0^{D-5} (r_0^2 + a_1^2)\,,\qquad J^0_1=\fft{2}{D-2}M_0 a_1\,,\qquad S_0 = \fft{4\pi}{D-2} M_0 r_0\,,\cr
&&T_0=\frac{(D-3)r_0^2+(D-5)a_1^2}{4\pi r_0(r_0^2+a_1^2)}\,,\qquad \Omega_0^1=\frac{a_1}{r_0^2+a_1^2}\,.
\eea
As was explained earlier, in this case, there can be no extremal limit, and the black hole must have non-vanishing temperature. Under the quadratic perturbation with fixed $(T_0, \Omega_0^1)$, the leading order corrections to the thermodynamic variables are
\bea\label{J4}
\Delta M &=& -\frac{\alpha  (D-4) (D-3) \Sigma_{D-2} r_0^{D-7}}{16 \pi  \left(a_1^2+r_0^2\right)}\Big((D-2)^2 r_0^4+2 a_1^2 \left(D^2-6 D+6\right) r_0^2\,,\nn\\
&&+ (D-4)^2a_1^4\Big),\nn\\
\Delta S &=& -\frac{\alpha  (D-3) \Sigma_{D-2} r_0^{D-6} }{4 \left(a_1^2+r_0^2\right) \left((D-3) r_0^2-a_1^2 (D-5)\right)}
\Big((D-5) (D-2)^2 r_0^6 -(D-7)) (D-4)^2a_1^6\nn\\
&&+ \left(D^3-11 D^2+32 D-24\right) a_1^2 r_0^4 - \left(D^3-13 D^2+48 D-44\right) a_1^4 r_0^2
\Big),\nn\\
\Delta J^1 &=& -\frac{\alpha (D-4) (D-3)\Sigma_{D-2}\, a_1 r_0^{D-7} }{8 \pi  \left(a_1^2+r_0^2\right) \left(a_1^2 (D-5)-(D-3) r_0^2\right)}
\Big(\left(-D^2+3 D-4\right) r_0^6\cr
 &&- \left(D^2-5 D-4\right)a_1^2 r_0^4+ (D-4) (D-3)a_1^4 r_0^2+ (D-5) (D-4)a_1^6\Big).
\eea
We see that as was understood there is no divergence in $\Delta M$ for all the parameters. However, both $\Delta S$ and $\Delta J^1$ become singular when we have $r=r_*$, with the critical radius
\be
r_*=\sqrt{\ft{D-5}{D-3}}\, a_1\,,\label{r0a1}
\ee
for which the temperature and angular velocity given by
\be
T_*=\frac{\sqrt{D-5} (D-3)^{3/2}}{4 \pi (D-4)a_1 }=\fft{\sqrt{(D-5)(D-3)}}{2\pi} \Omega_0^1\,.
\ee
Thus, we clearly see an example in all dimensions higher than 5 that the perturbative approach breaks down at certain specific non-zero temperature. This raises a question: how can we determine this temperature? The divergence is, in fact, already rooted in the original uncorrected thermodynamics. It follows from the perturbed first law \eqref{pertfl} that
\be
\Delta S = -\fft{\partial \Delta G}{\partial T_0}\Big|_{\Omega_0^1}\,,\qquad
\Delta J^1=-\fft{\partial \Delta G}{\partial \Omega^1_0}\Big|_{T_0}\,,
\ee
Note that $\Delta G=\Delta G(r_0,a_1), T_0=T_0(r_0,a_1)$ and we can express $a_1$ either as $a_1=a_1(r_0,\Omega_0^1)$ or as $a_1=a_1(r_0,T_0)$. Therefore, we have
\be
\Delta S = -\Big(\fft{\partial\Delta G}{\partial r_0}\Big|_{\Omega_0^1}\Big)\Big/
\Big(\fft{\partial T_0}{\partial r_0}\Big|_{\Omega_0^1}\Big),\qquad
\Delta J^1 = -\Big(\fft{\partial\Delta G}{\partial r_0}\Big|_{T_0}\Big)\Big/
\Big(\fft{\partial \Omega^1_0}{\partial r_0}\Big|_{T_0}\Big).
\ee
We see that the divergence is already encoded in the unperturbed thermodynamics, namely the zeros of the two quantities $\fft{\partial T_0}{\partial r_0}\Big|_{\Omega_0^1}$ and $\fft{\partial \Omega^1_0}{\partial r_0}\Big|_{T_0}$. It is straightforward to verify that we have
\be
\fft{\partial T_0}{\partial r_0}\Big|_{\Omega_0^1}= \{T_0,\Omega_0^1\}\Big/{\fft{\partial\Omega_0^1}{\partial a}\Big|_{r_0}}\,,\qquad
\fft{\partial \Omega^1_0}{\partial r_0}\Big|_{T_0}=\{\Omega_0^1,T_0\}\Big/{\fft{\partial T_0}{\partial a}\Big|_{r_0}}\,.
\ee
Here, $\{f,g\}$ is the Poisson bracket with respect to variables $(r_0,a_1)$. Thus the vanishing of
\be
\{T_0, \Omega_0^1\}=\frac{a^2 (D-5)-(D-3) r_0^2}{4 \pi  r_0^2 \left(a^2+r_0^2\right)^2}\,,
\ee
corresponding to \eqref{r0a1}, is precisely where the perturbation breaks down. It is of interest to note that now we have a simple way to calculate $\Delta S$ and $\Delta J_0^1$, given by
\be
\Delta S= -\fft{\{\Delta G, \Omega_0^1\}}{\{T_0, \Omega_0^1\}}\,,\qquad
\Delta J_0^1 =- \fft{\{\Delta G, T_0\}}{\{\Omega_0^1, T_0\}}\,.
\ee

However, for general configurations of the angular momenta, we cannot adopt the above Poisson bracket. Nevertheless, there is still a simple way to derive the critical point, which corresponds to having $dT_0$ and all $d\Omega_0^i$ vanish.  Introducing $X^\mu= (T_0, \Omega_0^i)$ and $x^\mu = (r_0, a_i)$, the critical point is located at the vanishing of the determinant of the $(N+1)\times (N+1)$ matrix $M_{\mu\nu}$, namely
\be
\det(M^\mu{}_\nu) = 0\,,\qquad M^{\mu}{}_{\nu} \equiv \fft{\partial X^\mu}{\partial x^\nu}\,.
\ee
For our thermodynamic variables \eqref{tsomega}, the above condition implies that the following factor $Z$ must vanish, namely
\be
\det(M_{\mu}{}_{\nu})\sim Z \equiv\fft{1}{\epsilon+1} - \sum_i \fft{r_*^2}{r_*^2 - a_i^2}=0\,.\label{breakdowncond}
\ee
Thus the critical temperature is
\be
T_* = \fft{r_0}{2\pi}\sum_{i=1}^N \fft{2 a_i^2}{a_i^4-r_*^4}\,.
\ee
That the critical temperature $T_0^*$ is positive is not obvious, and it is the indication that $r_*$ is indeed the event horizon of the black hole. This is because, as we discussed earlier, for given $(m, a_i)$ parameters, there can be  one positive horizon $r_+$ or at most two positive horizons $r_-\le r_+$, with $T_0(r_-)<0$ and $T_0(r_+)>0$.

Note that when $a_i$'s are all equal, then $X$ has no real roots. This is precisely the case we considered earlier, with no perturbative breakdown. However, the breakdown does occur for other general $a_i$ configurations. We have not been able to prove that the breakdown condition would always lead to $T_*\ge 0$, but we have tried many randomly selected $a_i$ to confirm this numerically. Here, we illustrate this with a concrete example. We consider $D=7$ and hence $N=3$.

As an example for illustration, we choose $a_i=i$. The condition $X=0$ gives two positive $r_*$, namely $r_*=2.5661$ and $r_*=1.3689$. Both corresponding $T_*$ are positive, indicating that $r_*=r_+$, i.e.~the event horizon. In table 1, we present the relevant data of the black holes for the two $r_*$'s. These two are perfectly well-defined rotating black holes, but do not accept the perturbation, which would diverge.

\begin{center}
\begin{tabular}{|c|c|c|c|c|}
  \hline
  $r_*$ & $T_*$ & $m$ & $r_+$ & $r_-$ \\ \hline
  2.5661 & 0.056612 & 95.008 & 2.5661 & 0.5121 \\ \hline
  1.3689 & 0.016671 & 48.979 & 1.3689 & 1.06294 \\
  \hline
\end{tabular}
\end{center}
{\small Table 1. The two perturbative breakdowns for given $(a_1,a_2,a_3)=(1,2,3)$ in $D=7$. In each case, we have $r_*=r_+$, the event horizon, which is consistent with $T_*>0$.}

\section{Corrections in the canonical ensemble}
\label{sec:F}

Thermodynamic potential of the canonical ensemble is the Helmholtz free energy, given by
\be
F=M - T S = G + \sum_i\Omega^i J^i\,.
\ee
We cannot directly calculate this free energy using the Euclidean action approach, but can via the above Legendre transformation after having obtained the Gibbs free energy. The corresponding first law is
\be
dF = - S dT + \sum_i \Omega^i dJ^i\,.
\ee
Thus we should keep the thermodynamic variables $(T,J^i)=(T_0, J_0^i)$ fixed under the quadratic curvature correction. This can be achieved by making appropriate redefinition of the parameters $(r_0, a_i)$ at the linear $\alpha$ order
\be
r_0\rightarrow r_0 + \alpha \Delta r_0\,,\qquad a_i\rightarrow a_i + \alpha \Delta a_i\,.\label{r0airedef}
\ee
In the grand canonical ensemble we studied in the previous section, we have obtained $T=T_0$ and $J^i= J^i_0 + \Delta J^i$, both as functions of $(r_0,a^i)$. Substituting \eqref{r0airedef} into $(T, J^i)$ and requiring $(T,J^i)=(T_0, J_0^i)$  at the linear $\alpha$ order, we can solve for $(\Delta r_0, \Delta a_i)$. We thus obtain the remaining (corrected) thermodynamic quantities. The first law of these corrected variables is
\be
d\Delta F = -\Delta S d T_0 + \sum_i \Delta \Omega^i dJ_0^i\,.\label{ffl}
\ee
It is easy to show that the generalized Smarr relation in this ensemble is
\be
(D-5) \Delta F = \Delta S\, T_0 + (D-4) \sum_{i} \Delta \Omega^i J_0^i\,.
\ee
The general expressions for all the corrected thermodynamic variables can be very complicated, and we present only some special cases.

\subsection{All equal angular momenta}

The system simplifies when all angular momenta are equal, with $a_i=a$ and $J=N J_i$. In odd dimensions, we have
\bea\label{O1}
\Delta M &=&\frac{\alpha  (D-3)\Sigma_{D-2} \left(a^2+r_0^2\right)^{\frac{D-5}{2}}}{16 \pi  r_0^4 \left(2 a^4 D+a^2 \left(D^2-3 D+6\right) r_0^2-(D-3) r_0^4\right)}\Big(
2 a^8 (D+2)\cr
&&+a^6 \left(5 D^2-47 D+66\right) r_0^2-a^4 \left(2 D^3-39 D^2+153 D-152\right) r_0^4\cr
&&+a^2 (D-4) \left(D^3-11 D^2+36 D-34\right) r_0^6+
(D-2)^2 \left(D-3\right)\left(D-4\right) r_0^8\Big),\cr
\Delta S&=& \frac{\alpha  (D-3) \Sigma_{D-2} \left(a^2+r_0^2\right)^{\frac{D-3}{2}}}{4 r_0^3 \left(2 a^4 D+a^2 \left(D^2-3 D+6\right) r_0^2-(D-3) r_0^4\right)}\Big(
-2 a^6 (D+1)\cr
&&+3 a^4 (7 D-11) r_0^2+2 a^2 \left(D^3-7 D^2+23 D-23\right) r_0^4
+(D-5) (D-2)^2 r_0^6\Big),\\
\Delta \Omega &=&
\frac{\alpha  (D-3)a \left(2 a^6-7 a^4 (D-3) r_0^2-2 a^2 \left(D^2-5 D+1\right) r_0^4+(D-4) \left(D^2-3 D+4\right) r_0^6\right)}{r_0^2 \left(a^2+r_0^2\right){}^2 \left(-2 a^4 D-a^2 \left(D^2-3 D+6\right) r_0^2+(D-3) r_0^4\right)}\,.\nn
\eea
We see that the critical divergent point $r_*$ is given by
\be
(D-3) r_*^4 - \left(D^2-3 D+6\right) a^2 r_*^2-2D a^4 =0\,.
\ee
In even dimensions, we have similar results
\bea\label{O2}
\Delta M &=& \frac{\alpha  (D-4) (D-3) (D-2) \Sigma_{D-2} \left(2 a^4-a^2 (D-4) r_0^2+(D-3) (D-2) r_0^4\right)}{16 \pi  r_0 \left(a^4 (D-1)+a^2 \left(D^2-4 D+6\right) r_0^2-(D-3) r_0^4\right)\left(a^2+r_0^2\right)^{\frac{4-D}{2}}}\,,\cr
%%%%
\Delta S &=& \frac{\alpha  (D-3) (D-2)\Sigma_{D-2}\left(6 a^4+a^2 \left(2 D^2-9 D+16\right) r_0^2+(D-5) (D-2) r_0^4\right)}{4 \left(a^4 (D-1)+a^2 \left(D^2-4 D+6\right) r_0^2-(D-3) r_0^4\right)\left(a^2+r_0^2\right)^{\frac{4-D}{2}} }\,,\cr
%%%
\Delta\Omega &=& \frac{\alpha  (D-4) (D-3) a\left(2 a^4+a^2 (D-2) r_0^2- \left(D^2-3 D+4\right) r_0^4\right)}{\left(a^2+r_0^2\right)^2 \left(a^4 (D-1)+a^2 \left(D^2-4 D+6\right) r_0^2-(D-3) r_0^4\right)}\,.
\eea
The critical divergent point $r_*$ is thus determined by
\be
(D-3) r_*^4 - \left(D^2-4 D+6\right) a^2 r_*^2 -(D-1)a^4=0\,.
\ee
Both examples gives non-vanishing critical temperature $T_*=T_0(r_*,a)>0$.

\subsection{Nearly all equal angular momenta}

In odd dimensions, there is a case of nearly all equal angular momenta, defined by \eqref{nearly}. We find
\bea\label{O3}
\Delta M &=& \frac{\alpha  (D-3) \Sigma_{D-2} \left(a^2+r_0^2\right)^{\frac{D-7}{2}}}{16 \pi  r_0^2 \left(r_0^2-a^2 (D-2)\right)}
\Big(3 a^6 (D-2)+a^4 (5 D-8) r_0^2\cr
&&-a^2 (D-4) \left(D^2-2 D+2\right) r_0^4-(D-4) (D-2)^2 r_0^6\Big).\cr
\Delta S &=& \frac{\alpha  \Sigma_{D-2}\left(a^2+r_0^2\right)^{\frac{D-5}{2}}}{4 r_0^3 \left(r_0^2-a^2 (D-2)\right)}\Big(2 a^6 (D-2)+3 a^4 (D-1) r_0^2\cr
&&-2 a^2 \left(D^3-6 D^2+11 D-9\right) r_0^4-(D-5) (D-2)^2 r_0^6
\Big),\cr
\Delta \Omega &=& \frac{a \alpha  \left((D-7) \left(a^4+2 a^2 r_0^2\right)+(D-4) \left(D^2-3 D+4\right) r_0^4\right)}{r_0^2 \left(a^2+r_0^2\right){}^2 \left(r_0^2-a^2 (D-2)\right)}\,,
\eea
The divergent critical $r_*$ is
\be
r_*=\sqrt{D-2} a\,,
\ee
for which $T_0(r_*)$ is some positive value. Note that in the grand canonical ensemble, the divergence occurs at zero temperature for this charge configuration.

\subsection{Single angular momentum}

Another simpler example is to consider single angular momentum. We have
\bea\label{O4}
\Delta M&=&- \frac{\alpha (D-3)(D-4)\Sigma_{D-2}r_0^{D-7}}{16\pi(a^2+r_0^2)(3(D-5)a^4-
6a^2r_0^2+(D-3)r_0^4)}\Big( (D - 8) (D - 5) (D - 4)a^8\cr
&&+2(2D^3-27D^2+106D-114)a^6r_0^2+2(3D^3-32D^2+98D-82)a^4r_0^4\cr
&&+2(2D^3-17D^2+46D-38)a^2r_0^6+(D-2)^2(D-3)r_0^8\Big),\cr
\Delta S &=& -\frac{ \alpha (D-3) \Sigma_{D-2} r_0^{D-6}}{4(3(D-5)a^4-6a^2r_0^2+(D-3)r_0^4)}\Big((D-11)(D-4)^2a^6+
(D-2)^2(D-5)r_0^6\cr
&&+3(D^3-13D^2+48D-44)a^4r_0^2+(3D^3-29D^2+96D-88)a^2r_0^4\Big),\nn\\
\Delta\Omega &=&\frac{\alpha (D-3)(D-4)a }{r_0^2(r_0^2+a^2)^2\left(3(D-5)a^4-6a^2r_0^2+(D-3)r_0^4\right)}
\Big((-D^2+9D-20)a^6\cr
&&-(D^2-7D+12)a^4r_0^2 +(D^2-5D-4)a^2r_0^4+(D^2-3D+4)r_0^6\Big).
\eea
The critical temperature $T_0(r_*)$ is determined by
\be
(D-3)r_*^4-6a^2r_*^2+3(D-5)a^4=0\,.
\ee

\subsection{General case}

In all above examples, we have effectively two independent parameters, $(r_0, a)$. We can use the Poisson bracket method to calculate all the corrected thermodynamic quantities. For the general case with $(r_0, a_i)$, the critical temperature can be determined by the $(N+1)\times (N+1)$ matrix, defined as follows. We set $Y^{\mu}=(T_0,J^i_0)$ and $x^\mu=(r_0,a_i)$, and define the matrix
\be
		N_{\mu\nu}\equiv\frac{\partial Y^\mu}{\partial x^\nu}\,.
\ee
The critical temperature of divergence is determined by
\be
\det(N_{\mu\nu})=0\,,\qquad\rightarrow\qquad \frac{D-2}{D-2-4\pi r_0 T_0}\sum_{i}^{N}\frac{a_i^2}{(r_0^2+a_i^2)^2}=\fft{\pi T_0}{r_0}\,.\label{cancons}
\ee
Thus we see that zero temperature $T_0=0$ cannot be a solution of the above equation. Furthermore, if there is a real solution $r_0=r_*$, then $T_0(r_*)$ must be positive, indicating that the corresponding $r_*$ is indeed the outmost horizon, i.e.~the event horizon. While we cannot prove that \eqref{cancons} must have a real solution for given set of $a_i$'s, the explicitly examples we have already discussed illustrate that such a real solution generally exists, giving rise the critical point where the linear perturbation approach breaks down.

\section{Corrections with fixed mass and angular momenta}
\label{sec:MJ}

\subsection{Corrections and zero-temperature divergence}

It is natural to consider higher-order curvature corrections to black holes with fixed conserved quantities, such as mass and angular momenta. The thermodynamic potential for such a system is the entropy, with the first law
\be
dS = \fft{1}{T} dM + \sum_{i}\fft{\Omega_i}{T} dJ^i\,.
\ee
With $(M,J^i)=(M_0, J_0^i)$ fixed under the perturbation, the first law for the remainder corrected thermodynamic variables is
\be
   	\Delta T dS_0+T_0 d\Delta S+\sum_i \Delta \Omega_{i}dJ^i_0=0\,.\label{mjfixfl}
\ee
This implies a Smarr-like relation
\be
(D-2) \Big(\Delta T S_0 + \sum_i\Delta\Omega_i J_0^i\Big) + (D-4) T_0 \Delta S=0\,.
\label{mjfixsmarr}
\ee
The divergent critical temperature is at vanishing temperature, $T_*=0$. This can be seen from the theorem proved in \cite{Reall:2019sah} that for fixed mass and angular momenta, one has
\be
\Delta S= -\Delta I = \fft{\Delta G}{T_0}\,,
\ee
$\Delta G$, given in \eqref{DeltaG}, are regular for all $(r_0, a_i)$. Thus we see that $\Delta S$ is singular at zero $T_0$. Substituting the above into \eqref{mjfixfl}, we have
\be
\Delta T dS_0 + \sum_i \Delta \Omega_i \Delta J_0^i = \fft{\Delta G}{T_0} dT_0 - d\Delta G\,.
\ee
This, together with \eqref{mjfixsmarr}, implies that $\Delta T$ and $\Delta \Omega^i$ are also singular at $T_0=0$. This breakdown was also observed in
\cite{Cheung:2018cwt,Horowitz:2024dch}, where the origin was understood that for fixed mass and charge, zero temperature cannot be reached. It was suggested \cite{Horowitz:2024dch} that the divergence will disappear for fixed temperature. The fact that fixed mass and charge cannot reach zero temperature in itself does not necessarily imply that it has to diverge, and we had the $D=5$ example in section \ref{sec:G} where the correction is divergent even for the case at fixed zero temperature.

As we have discussed in section \ref{sec:mp}, except for all equal angular momenta or nearly all equal angular momenta cases, extremal limits do not exist and hence the perturbative breakdown will not occur for fixed mass and angular momenta. For example, for the single angular momentum case, we have
\bea\label{T4}
\Delta\Omega &=& -\frac{2\alpha (D-4) (D-3) a}{(D-2) r_0^2 \left(a^2+r_0^2\right)^3
			\left(a^2 (-5+D)+(-3+D) r_0^2\right)}\Big((D-4) (D-5) a^6\cr
&&+\left(10-11 D+2 D^2\right) a^4
			r_0^2+D (D-3) a^2 r_0^4-(D-2) r_0^6\Big)\,,\nn \\
\Delta T &=& -\frac{\alpha (D-3)(D-4) }{4 (D-2) \pi  r_0^3 (a^2+r_0^2)^3 \left(a^2 (D-5)+(D-3) r_0^2\right)}\Big(\left( (D-8) (D-5) (D-4)a^8\right.\cr
&&\left.+2  (-114+D (106+D (2D-27)))a^6 r_0^2+2  (-82+D (98+D (3D-32)))a^4 r_0^4\right.\cr
&&\left.+2  (-38+D (46+D (2D-17)))a^2 r_0^6+(D-3)
		(D-2)^2 r_0^8\right)\Big)\cr
\Delta S &=&\frac{\alpha (D-3)\Sigma_{D-2} r_0^{D-6}}{4(a^2 (D-5)+(D-3) r_0^2)}\Big((D-4)^2a^4+2\left(6-6 D+D^2\right)
			a^2 r_0^2+(D-2)^2 r_0^4\Big)\,.
\eea
For $D=4,5$, the $T_0=0$ extremal limit exists, where divergence occurs, but as we expected, there is no divergence for $D\ge 6$.

For all equal angular momentum, in odd dimensions, we have
\bea\label{T1}
\Delta T &=& -\frac{\alpha  (D-3)}{4 \pi  (D-2) r_0^3 \left(a^2+r_0^2\right)^3 \left((D-3) r_0^2-2 a^2\right)}\Big(2(D+2)a^8\cr
&&+\left(5 D^2-47 D+66\right) a^6 r_0^2 -\left(2 D^3-39 D^2+153 D-152\right) a^4 r_0^4 \cr
&&+ (D-4) \left(D^3-11 D^2+36 D-34\right)a^2 r_0^6 +(D-4) (D-3) (D-2)^2 r_0^8
\Big),\nn\\
\Delta S &=& \frac{\alpha  (D-3) \Sigma_{D-2} \left(a^2+r_0^2\right)^{\frac{D-3}{2}} \left(a^4-2  (2 D-3) a^2 r_0^2+(D-2)^2 r_0^4\right)}{4 r_0^3 \left((D-3) r_0^2-2 a^2\right)}\,,\nn\\
\Delta\Omega &=& \frac{2 \alpha  (D-3) a \left(2 a^6+ (1-2 D) a^4 r_0^2+2 (D-3) (D-1) a^2 r_0^4+(D-4) (D-2) r_0^6\right)}{(D-2) r_0^2 \left(a^2+r_0^2\right)^3 \left((D-3) r_0^2-2 a^2\right)}\,.\label{odddelta}
\eea
The expressions in even dimensions are simpler
\bea\label{T2}
\Delta T &=& -\frac{\alpha  (D-4) (D-3) \left(2 a^4- (D-4) a^2 r_0^2+(D-3) (D-2) r_0^4\right)}{4 \pi  r_0 \left(a^2+r_0^2\right){}^2 \left((D-3) r_0^2-a^2\right)}\,,\nn\\
\Delta S &=& \frac{\alpha  (D-3) (D-2) \Sigma_{D-2} \left(a^2+r_0^2\right){}^{\frac{D}{2}-2} \left((D-2) r_0^2-2 a^2\right)}{4 \left((D-3) r_0^2-a^2\right)}\,,\nn\\
\Delta\Omega &=& \frac{2\alpha  (D-4) (D-3)a r_0^2}{\left(a^2+r_0^2\right){}^2 \left((D-3) r_0^2-a^2\right)}\,.\label{evendelta}
\eea
In both cases, as we have expected, the critical temperature for divergence is zero. For the nearly all equal angular momentum case, we have
\bea\label{T3}
\Delta T &=& \frac{\alpha  (D-3)}{4 \pi  (D-2) r_0^3 \left(a^2+r_0^2\right)^3}
\Big(3(D-2)a^6+ (5 D-8)a^4 r_0^2\cr
&&- (D-4) \left(D^2-2 D+2\right) a^2 r_0^4-(D-4) (D-2)^2 r_0^6\Big),\nn\\
\Delta S &=& \frac{\alpha \Sigma_{D-2}}{4 r_0^3} \left(a^2+r_0^2\right)^{\frac{D-5}{2}} \left(a^4+2 a^2 r_0^2+(D-2)^2 r_0^4\right)\,,\nn\\
\Delta \Omega &=& -\frac{2 a \alpha  \left(a^4 (2 D-5)+2 a^2 (2 D-5) r_0^2-(D-4) (D-2) r_0^4\right)}{(D-2) r_0^2 \left(a^2+r_0^2\right)^3}\,.
\eea

\subsection{On the sign of $\Delta S$}

It was pointed out that for the system with fixed mass, angular momenta and charges, one should expect $\Delta S>0$ for stable thermodynamics \cite{Cheung:2018cwt}. We have already pointed it out earlier that Myers-Perry black holes have unstable thermodynamics for all allowed parameters, if we consider mass and all angular momenta as thermodynamic variables. In black hole thermodynamics, stability is rather rare. Even for the Reissner-Nordstr\"om (RN) black hole, its thermodynamics cannot be stable if both mass and charge are treated as thermodynamic variables, since the product of the specific heat capacity and charge capacitance is always negative \cite{Ma:2020xwi}. One way to resolve this instability issue for the RN case is to fix the charge to a specific value so that it does not involve in the thermodynamic process, in which case, the specific heat capacity with fixed $Q$ is positive for low temperature. We can follow the same strategy, and fix all the angular momenta to some appropriate specific values to extract them from the thermodynamic process. As an example, for simplicity, we consider the case with all equal angular momenta $J_i=J/N$. For odd dimensions with $D=2N+1$, we find that the specific heat capacity is
\be
C_J = \fft{\partial M_0}{\partial T_0}\Big|_J = \frac{\pi  (D-2) \Sigma_{D-2}  \left(a^2+r_0^2\right)^{\frac{D+3}{2}} T_0}{2D a^4+ \left(D^2-3 D+6\right) a^2 r_0^2-(D-3) r_0^4}\,.
\ee
We see that for low temperature, or equivalently smaller $r_0$, $C_J$ is positive. The thermodynamics is stable when $r_0$ lies in the range
\be
\frac{2 a^2}{D-3}\equiv r_{\dagger}^2\le r_0^2 \le r_{\sharp}^2 \equiv \frac{ D^2-3 D+6+\sqrt{(D-2) (D-1) \left(D^2-3 D+18\right)}}{2 (D-3)} a^2\,.
\ee
Here $r_\dagger$ is the radius parameter of the extremal black hole $(T_0=0)$. Beyond $r_\sharp$, the capacity is negative and the thermodynamics becomes unstable. On the other hand, it follows from \eqref{odddelta}, the sign of $\Delta S/\alpha$ in the full ranges of thermodynamic stability is given by
\bea
\fft{\Delta S}{\alpha} <0:&&\quad \frac{2 a^2}{D-3} \le r_0^2 < r_{\ddagger}^2 \equiv \frac{(2 D-3)+\sqrt{(D-1)(3 D-5)}}{(D-2)^2} a^2 <r_\sharp^2\,,\cr
\fft{\Delta S}{\alpha} >0:&&\quad r_{\ddagger}^2 <r_0^2 \le r_\sharp^2\,.
\eea
Thus we see that $\Delta S$ cannot be always positive in the stability region regardless the sign choice of the coupling constant $\alpha$.

In even $D=2N+2$ dimensions with all equal angular momenta, the story is similar. We have the specific heat capacity
\be
C_J=\frac{\pi  (D-2)\Sigma_{D-2}   r_0  \left(a^2+r_0^2\right)^{\frac{D}{2}+1}T_0}{a^4 (D-1)+ \left(D^2-4 D+6\right)a^2 r_0^2-(D-3) r_0^4}\,,
\ee
which implies that the the thermodynamic is stable in the following parameter region:
\be
\frac{ a^2}{D-3}\equiv r_\dagger^2 \le r_0^2\le r_\sharp^2\equiv\frac{6-4 D+D^2+(-2+D) \sqrt{12-4 D+D^2}}{2 (D-3)} a^2\,.
\ee
It follows from \eqref{evendelta} that $\Delta S$ can change its sign when $r_0$ crosses over $r_{\ddagger}$, with
\be
r_\dagger^2 < r_\ddagger^2 \equiv \frac{2 a^2}{D-2}<r_\sharp^2\,.
\ee
We thus find counterexamples of \cite{Cheung:2018cwt}, although it can be argued that it is much more difficult to fix the angular momenta than the electric charges in a thermodynamic process. A counterexample to \cite{Cheung:2018cwt} with varying angular momenta was found in the higher-derivative correction in supergravity \cite{Ma:2022gtm}.

\section{Kerr black hole under cubic curvature correction}
\label{sec:D4}

The quadratic curvature correction in four dimensions is trivial, we thus consider the cubic corrections for this special dimensions. There are two cubic Riemann tensor polynomials, namely
\bea
\Delta L=\sqrt{-g}\left( R+\beta R^{ijkl}R_{ji}{}^{mn}R_{nmlk}+\gamma R^{ijkl}R^{m}{}_{i}{}^{n}{}_{k}R_{jmln}\right).
\eea
In four dimensions, however, the cubic Lovelock combination, corresponding to setting $2\beta=\gamma$, vanishes identically. Thus, there is only one nontrivial cubic curvature invariant at the perturbative order in four dimensions, namely $\alpha = 2\beta -\gamma\ne 0$. The cubic correction to the thermodynamic of the Kerr metric was studied in \cite{Reall:2019sah}, but we here have a different emphasis. For the grand conical ensemble with fixed $(T,\Omega)$ under the perturbation, we have
\bea
T&=& T_0=\frac{r_0^2-a^2}{4\pi r_0(r_0^2+a^2)}\,,\qquad \Omega =\Omega_0=\frac{a}{r_0^2+a^2}\,,\nn \\
M &=& \frac{r_0^2+a^2}{2r_0} +\frac{\alpha \left(a^6-21 a^4 r_0^2+35 a^2 r_0^4-7 r_0^6\right)}{14 r_0^3 \left(a^2+r_0^2\right)^3}\,,\nn\\
S &=& \pi(r_0^2+a^2)-\frac{\alpha \pi \left(3 a^6-21 a^4 r_0^2-35 a^2 r_0^4+21 r_0^6\right)}{7 r_0^2 \left(a^2+r_0^2\right)^3}\,,\nn\\
J &=& \frac{r_0^2+a^2}{2r_0}a-\frac{\alpha a \left(9 a^4-14 a^2 r_0^2-7 r_0^4\right)}{7 r_0 \left(a^2+r_0^2\right)^3}\,.
\eea
In the canonical ensemble with fixed $(T_0, J_0)$, the corrected thermodynamic variables are
\bea
\Delta \Omega &=& -\frac{2\alpha a \left(9 a^4-14 a^2 r_0^2-7 r_0^4\right)}{7 \left(a^2+r_0^2\right)^3 \left(3 a^4+6 a^2 r_0^2-r_0^4\right)}\,,\nn\\
\Delta M &=&\frac{\alpha \left(3 a^{10}-21 a^8 r_0^2-42 a^6 r_0^4+126 a^4 r_0^6-105 a^2 r_0^8+7 r_0^{10}\right)}{14 r_0^3 \left(a^2+r_0^2\right)^3 \left(3 a^4+6 a^2r_0^2-r_0^4\right)}\,,\nn\\
\Delta S &=&\frac{\alpha \pi \left(-9 a^8+90 a^6 r_0^2+196 a^4 r_0^4-266 a^2 r_0^6+21 r_0^8\right)}{7 r_0^2 \left(a^2+r_0^2\right)^2 \left(3 a^4+6 a^2r_0^2-r_0^4\right)}\,.
\eea
For fixed $(M, J)=(M_0,J_0)$, we have
\bea
\Delta\Omega &=& \frac{\alpha \left(a^9-38 a^5 r_0^4+56 a^3 r_0^6-35 a r_0^8\right)}{7 r_0^2 \left(-a^2+r_0^2\right) \left(a^2+r_0^2\right)^5}\,,
\quad \Delta S =\frac{\alpha \pi \left(a^6-21 a^4 r_0^2+35 a^2 r_0^4-7 r_0^6\right)}{7 r_0^2 \left(-a^2+r_0^2\right) \left(a^2+r_0^2\right)^2}\,.\nn\\
\Delta T &=&-\frac{\alpha \left(3 a^{10}-21 a^8 r_0^2-42 a^6 r_0^4+126 a^4 r_0^6-105 a^2 r_0^8+7 r_0^{10}\right)}{28 \pi  r_0^3 \left(-a^2+r_0^2\right)
    		\left(a^2+r_0^2\right)^5}\,.
\eea
In all these examples, whether the corrected thermodynamic quantities are divergent or not depends on the original uncorrected thermodynamic variables, as was discussed in higher dimensional cases. The cubic extension follows the same rule.

It is worth pointing out again that the specific heat capacity is
\be
C_J=\frac{8 \pi ^2 r_0 \left(a^2+r_0^2\right){}^3}{3 a^4+6 a^2 r_0^2-r_0^4}\,.
\ee
We therefore have stable thermodynamics for the fixed angular momentum in the following parameter region.
\be
a^2\le r_0^2\le r_\sharp^2\equiv\left( 3+2 \sqrt{3}\right) a^2\,.
\ee
However $\Delta S$ is not positive definite in the stable region. $\Delta S$ charges its sign when $r_0$ cross over $r_\ddagger\sim 2.077a$.

\section{Divergence is a perturbative artefact: an explicit example}
\label{sec:artefact}

In the previous sections, we considered quadratic curvature corrections to thermodynamic quantities of Myers-Perry black holes in diverse dimensions. We found that for different thermodynamic ensembles, there are all inevitable divergences for generic parameters that break down the perturbative approach. It is natural to ask the question: are these breakdowns real or the artefact of perturbation? There are plenty of examples of the linear perturbative conclusions that do not survive the full nonlinear level. For example, AdS spacetime is stable under the linear perturbation, but unstable at the nonlinear level \cite{Bizon:2011gg}. The horizon of higher-dimensional rotating black hole at the extremal limit can be destroyed perturbatively \cite{Mao:2023qxq}. Extremal charged black hole in Einstein-Born-Infeld gravity is superradiantly unstable no matter how small the Born-Infeld coupling is, although perturbatively it is stable \cite{Wu:2024fvy}. It is thus necessary to examine our perturbative divergence at the full nonlinear level.

There exist exact solutions of spherically-symmetric and static black holes in Einstein gravity extended with higher-derivative corrections. A notable example is charged black hole in Einstein-Maxwell gravity extended with a Gauss-Bonnet term, which is nontrivial only in $D\ge 5$. To illustrate our point, it is suffice to consider only $D=5$. The generalized RN black hole is given by the metric profile function \cite{Wiltshire:1985us}
\be
f=1+\frac{r^2 }{2 \alpha }\left(1-\sqrt{1+\frac{8 \alpha  \mu }{r^4}-\frac{4 \alpha  q^2}{r^6}}\right).\label{gbf}
\ee
The solution is parametrized by integration constants $(\mu,q)$, associated with the mass and electric charge
\be
M=\fft{3\pi}{4} \mu\,,\qquad Q_e=\fft{\sqrt3}{4}\pi q\,.
\ee
For simplicity, we shall simply call $(\mu,q)$ as the mass and charge respectively. The remainder of thermodynamic quantities $(T,S,\Phi_e)$ are
\be
T=\frac{r_0^4-q^2}{2 \pi  r_0^3 \left(2 \alpha +r_0^2\right)}\,,\qquad S=\frac{1}{2} \pi ^2 r_0 \left(r_0^2+ 6 \alpha \right)\,,\qquad \Phi_e=\frac{\sqrt{3} q}{r_0^2}\,,
\ee
where horizon $r_0$ is understood to be the largest root of its rational polynomial relation to the mass and charge:
\be
\mu=\frac{1}{2 r_0^2}\left(r_0^4 + q^2+\alpha  r_0^2\right)\,.
\ee
We now consider the case that the conserved quantities, the mass and charge, are fixed. We can express $r_0$, the largest root as a function of $(\mu,q)$:
\be
r_0=\sqrt{\ft{1}{2} \left(2 \mu-\alpha +\sqrt{(2 \mu-\alpha )^2-4 q^2}\right)}\,.
\label{r0muq}
\ee
We thus have
\bea
T &=&\frac{\sqrt{2} y}{\pi  \sqrt{2 \mu-\alpha +y} (2 \mu+3 \alpha +y)}\,,\qquad
S = \frac{\pi ^2 \sqrt{2 \mu-\alpha +y} (2 \mu +11 \alpha +y)}{4 \sqrt{2}}\,,\cr
\Phi_e&=& \frac{2 \sqrt{3} q}{2 \mu-\alpha +y}\,,\qquad y=\sqrt{(2\mu-\alpha)^2 - 4q^2}\,.
\eea
The extremal limit corresponds to taking $\mu=\mu_{\rm ext}$ with
\be
\mu_{\rm ext} = q + \ft12 \alpha\,.
\ee
We thus see that the thermodynamic variables are well behaved for all $\mu\ge \mu_{\rm ext}$. Note that for $\alpha=0$,  the extremal limit is $\mu_{\rm ext}^0=q$. For positive $\alpha$, $\mu$ can exceed $\mu_{\rm ext}^0$; for negative $\alpha$, it can not reach $\mu_{\rm ext}^0$. In either cases, there is no divergence in the vicinity of $\mu_{\rm ext}^0$ even for small $\alpha$.

We now pretend that there is no such full nonlinear exact solution and can only study these thermodynamic variables perturbatively order by order in $\alpha$. For fixed $(\mu,q)$, we have
\bea
T&=&\frac{\sqrt{\mu ^2-q^2}}{\pi  \left(\mu +\sqrt{\mu ^2-q^2}\right)^{3/2}}+\frac{\alpha  \left(\mu  \left(\sqrt{\mu ^2-q^2}-7 \mu \right)+5 q^2\right)}{4 \pi  \sqrt{\mu ^2-q^2} \left(\mu +\sqrt{\mu ^2-q^2}\right)^{5/2}}+O\left(\alpha ^2\right)\,,\nn\\
S &=&\frac{1}{2} \pi ^2 \left(\mu +\sqrt{\mu ^2-q^2}\right)^{3/2}+\frac{3 \pi ^2 \alpha  \sqrt{\mu +\sqrt{\mu ^2-q^2}} \left(7 \sqrt{\mu ^2-q^2}-\mu \right)}{8 \sqrt{\mu ^2-q^2}}+O\left(\alpha ^2\right)\,,\nn\\
\Phi_e &=& \frac{\sqrt{3} q}{\mu +\sqrt{\mu^2-q^2}}+\frac{\sqrt{3} \alpha  q}{2 \sqrt{\mu^2-q^2}\,(\sqrt{\mu^2-q^2}+\mu)}+O\left(\alpha ^2\right)\,.
\eea
Thus we see that at the linear $\alpha$'th order, the correction is divergent at the zero temperature, corresponding to $\mu_{\rm ext}^0=q$. It can be easily seen from the full nonlinear result that this divergence occurs at all perturbative orders of $\alpha$.

The origin of this divergence can also be understood at the level of the metric function \eqref{gbf}, which is a smooth function from the asymptotic infinity to the outer horizon, given by \eqref{r0muq}. Now if we use the perturbative approach, for fixed $(\mu,q)$, up to and including the $\alpha$ order, we have
\be
f=1 - \fft{2\mu}{r^2} + \fft{q^2}{r^4} + \alpha \frac{\left(q^2-2 \mu  r^2\right)^2}{r^{10}} + {\cal O}(\alpha^2)\,.
\ee
The outer horizon location is thus given by
\be
r_0 = \sqrt{\mu +\sqrt{\mu ^2-q^2}}\Big(1-\frac{\alpha }{4 \sqrt{\mu ^2-q^2}} +
{\cal O}(\alpha^2)\Big).
\ee
This clearly shows that the perturbative approach breaks down in the extremal or near extremal $(\mu-q)\sim \alpha$ region, while the full nonlinear result \eqref{r0muq} is perfectly regular.

While we do not have exact solutions of rotating black holes in Einstein gravity extended with quadratic or more general higher-order derivative terms, we expect that the divergence we explored in this paper is an artefact of perturbation, indicating a breakdown only of the perturbative approach. The divergence may not necessarily exist in the full nonlinear level.

\section{Conclusions}
\label{sec:con}

In this paper, we began a review of Myers-Perry rotating black holes in diverse dimensions. We presented a simple analytic proof that the rotating black holes in even dimensions are all thermodynamically unstable. Specifically, we showed that the trace of the corresponding thermodynamic metric has an elegant expression  \eqref{traceevend} and it is manifestly non-positive. However, for odd dimensions, we have to employ numerical analysis to prove the instability. We also found that the specific heat capacity of fixed angular momenta was positive at low temperature, including the extremal limit. This implies that the thermodynamics could be stable for the reduced system where all the angular momenta are fixed to some appropriate specific values and hence extracted from the thermodynamic process. We also found that in odd dimensions when one and only one angular momentum vanishes, there existed an intriguing decoupling limit of extremal solutions whose geometry \eqref{ads3} has an unusual AdS$_3$ embedding.

Our main focus was to carry out the derivation of the quadratic curvature corrections to the black hole thermodynamics of Myers-Perry solutions in general dimensions. We obtained an elegant result \eqref{DeltaG}, which allowed us to discuss various properties of black hole thermodynamics under the higher-derivative correction. We focused on one particular aspect of this perturbative approach: when does it break down? By breakdown, we mean that the correction becomes divergent so that no matter how small the coupling constant $\alpha$ is, the results cannot be trusted.

Intriguingly, in the grand canonical ensemble, when the angular momenta are all equal, there is no such breakdown. However, for generic parameters, or in other ensembles, breakdowns are common occurrences. We find that in this perturbative approach, the breakdowns are fully determined by the thermodynamics of the original uncorrected variables. Specifically, we assume that in a certain thermodynamic system, such as the canonical or grand canonical ensembles, the original uncorrected thermodynamic variables are $X^\mu$, which are functions of parameters $x^\mu$, (with $x^0=r_0$,) we can define a matrix $M^{\mu\nu}$ by
\be
dX^\mu = M^{\mu}{}_{\nu} dx^\nu\,.
\ee
The breakdown then could occur at $\det (M^\mu{}_\nu)=0$, since this quantity appears in the denominator of the corrected thermodynamic quantities. This provides a convenient early warning of the perturbative approach. It should be emphasized that this is only a necessary condition, but not sufficient. As we discussed in section \ref{sec:G}, there existed a rare example in the grand canonical ensemble that the numerator would cancel this zero precisely when all the angular momenta were nonzero and equal, thereby avoiding the breakdown.
Applying this condition to the Myers-Perry black holes, we have the condition \eqref{breakdowncond} for the grand canonical ensemble, and \eqref{cancons} for the canonical ensemble. It follows from \eqref{cancons} that the critical temperature $T_*$ can never be zero in the canonical ensemble, whilst breakdown always occurs at zero temperature for the system with fixed mass and angular momenta. For the grand canonical ensembles, critical temperature can be sometimes zero in some special cases, but it is nonzero in general.

In addition to the divergence of the $\alpha$-order correction, there are points in the parameter space  that the correction vanishes. What is intriguing is that these zeros occur at the same points for the corrections to different thermodynamic variables in different ensembles. For example, $\Delta J$ in grand canonical ensemble vanishes at the exact same point as $\Delta \Omega$ in canonical ensemble. This can be seen from the numerator of $\Delta J$ in eqns \eqref{J1}, \eqref{J2}, \eqref{J3}, \eqref{J4}, and compare them to the corresponding $\Delta \Omega$ in eqns \eqref{O1}, \eqref{O2}, \eqref{O3}, \eqref{O4}. The similar relation exists between $\Delta M$ in canonical ensemble and $\Delta T$ in system with fixed mass and angular momenta, as we can see from eqns \eqref{O1}, \eqref{O2}, \eqref{O3}, \eqref{O4} and \eqref{T1}, \eqref{T2}, \eqref{T3}, \eqref{T4}. These phenomena can be understood from the identities of differential equations, namely
\bea
 \left( \frac{\partial M}{\partial T}\Big|_{\alpha,J}\right)\,\left( \frac{\partial T}{\partial \alpha}\Big|_{M,J}\right) & = &-\left( \frac{\partial M}{\partial \alpha}\Big|_{T,J}\right) \,,\nn\\
 \,\left( \frac{\partial J}{\partial \Omega}\Big|_{\alpha,T}\right)\left( \frac{\partial \Omega}{\partial \alpha}\Big|_{T,J}\right) & = & -\left( \frac{\partial J}{\partial \alpha}\Big|_{\Omega,T}\right) \,.
\eea
Setting $\alpha=0$, we have
\bea
	C_{J_0}\,  \Delta T\Big|_{M_0,J_0} &=& -\Delta M\Big|_{T_0,J_0}\,,\qquad
	C_{J_0}=\left( \frac{\partial M_0}{\partial T_0}\Big|_{J_0}\right)\,,\nn\\
	I_{T_0}\,  \Delta \Omega\Big|_{T_0,J_0} &=& -\Delta J\Big|_{T_0,\Omega_0}\,,\qquad
	I_{T_0}=\left( \frac{\partial J_0}{\partial \Omega_0}\Big|_{T_0}\right)\,,\nn
\eea
where $C_{J_0}$ is the specific heat capacity  and $I_{T_0}$ is the isothermal differential moment of inertia.

We also examined the issue of the sign of $\Delta S$ under the quadratic curvature extension. Since there is only one nontrivial coupling constant $\alpha$, this would provide a strong condition on the sign choice of the $\alpha$. However, for the reduce stable thermodynamic system where all the angular momenta are taken to be some specific values, we found that $\Delta S$ could change sign within the stable parameter region, therefore providing counterexamples of \cite{Cheung:2018cwt}.

Finally, we used the exact solution of charged black hole in Einstein-Maxwell gravity extended with the Gauss-Bonnet term to study the perturbative divergence in the context of full nonlinear level. We showed that the divergence was an artefact of perturbation. In the full nonlinear solution, there is no such divergence. In this specific example, the Gauss-Bonnet term is part of the classical action, with no further higher-derivative corrections. This is different from the effective field theory approach, where the Gauss-Bonnet term is just the leading-order correction of an infinite series of higher-order terms. Regardless the situations, one needs to be cautious not to generalize the perturbative conclusion to the full nonlinear theory.

\section*{Acknowledgement}

We are grateful to Liang Ma and Yi Pang for discussions. This work is supported in part by the National Natural Science Foundation of China (NSFC) grants No.~11935009 and No.~12375052.

\appendix

\section{All thermodynamic quantities in grand canonical ensemble}
\label{app:thermo}

In section \ref{sec:G}, we adopted the RS method to obtain the general expression of the quadratic curvature tensor correction to Gibbs free energy, $\Delta G$ in \eqref{DeltaG}, of the grand canonical ensemble for general Myers-Perry black holes, where the thermodynamic variables are temperature and angular velocities. Based on some generalized Smarr-like relations, we showed that the correction to the mass is proportional to $\Delta G$, simply given by \eqref{granddeltam}. In this approach, the thermodynamic variables $(T_0, \Omega_0^i)$ remains fixed or uncorrected. The corrections occur to the other thermodynamic quantities, namely the entropy and angular momenta, and they can be deduced from the corrections to the first law \eqref{pertfl}, namely
\be
\Delta S = \fft{\partial \Delta G}{\partial T_0}\Big|_{\Omega_0^j}\,,\qquad
\Delta J^i=\fft{\partial \Delta G}{\partial \Omega_0^i}\Big|_{T_0, \Omega_0^{j\ne i}}\,.\label{deltaSJ}
\ee
In section 3, many explicit examples of these corrections were given, but only for some specific cases, such as single angular momentum, or all and nearly all equal angular momenta, {\it etc.} In this appendix, we give these quantities in its full generality.

We find it is useful to introduce notations ``dot'' and ``prime'', acting on a generic function $X(r_0,a_i)$ as follows
\be
\dot X = \fft{\partial X}{\partial r_0} \Big|_{\Omega_0^i}\,,\qquad
X' = \fft{\partial X}{\partial r_0}\Big|_{T_0, \Omega_0^{i\ne 1}}\,.
\ee
Specifically, we require that $\dot \Omega_0^i=0$, i.e.
\be
d\Omega_0^i = d\left( \frac{a_i}{r_0^2+a_i^2}\right) =0\,,\qquad\rightarrow
\qquad \dot{a_i}=\frac{2r_0 a_i}{r_0^2-a_i^2}\,.\label{dota}
\ee
Then the evaluation of $dX$ becomes
\be
dX = \fft{\partial X}{\partial r_0} dr_0 +
\sum_i \fft{\partial X}{\partial a_i} da_i =
\left(\fft{\partial X}{\partial r_0} + \sum_i \dot a_i\fft{\partial X}{\partial a_i} \right) dr_0 \equiv \dot X dr_0\,.
\ee
The $X'$ follows the analogous definition. With these definitions, the expressions of the corrections to the entropy and angular momenta in \eqref{deltaSJ} become
\be
\Delta S=-\frac{\dot{ \Delta G}}{\dot{ T_0}};\qquad
\Delta J^1=-\frac{\Delta G^\prime}{\Omega_0^{1\prime}},\qquad
\Delta J^{j\ne 1}=\Delta J^1\Big|_{a_1\leftrightarrow a_j}.\label{cSJ}
\ee
For convenience, we rewrite the expression of $\Delta G$ of \eqref{DeltaG} as:
\be
\Delta G=(-1)^{N}\frac{(D-3)\Sigma_{D-2}}{4\pi} \alpha\, m^2  \sum_{i}^{N}\frac{f_i}{b_i}\,,\qquad
f_i=\left\{
\begin{aligned}
	&\ft{r_0^4-6a_i^2r_0^2+a_i^4}{(r_0^2+a_i^2)^3}\,,\quad &D=\hbox{odd};\\
	&\ft{4r_0\left(r_0^2-a_i^2 \right) }{(r_0^2+a_i^2)^3}\,,\quad &D=\hbox{even},
\end{aligned}
\right.\label{NG}
\ee
and keep in mind that $b_i=\prod\limits_{j\neq i}(a_i^2-a_j^2)$. For this new expression \eqref{NG}, we have
\bea
\dot{\Delta G}&=&(-1)^{N+1}\frac{(D-3) \Sigma_{D-2}}{4\pi} \alpha \sum_{i}^{N}
\fft{f_i}{b_i}\Big(2m \dot{m}+m^2\big(\frac{\dot{f_i}}{f_i}-\frac{ \dot{b_i}}{b_i}\big)\Big),\label{dotG}\\
\Delta G^\prime&=&(-1)^{N+1}\frac{(D-3) \Sigma_{D-2}}{4\pi} \alpha \sum_{i}^{N}\fft{f_i}{b_i} \Big(2m m^\prime+m^2\big(\frac{f_i^\prime}{f_i}-\frac{ b_i^\prime}{b_i}\big)\Big).\label{primeG}
\eea
It follows from the expressions for $m$ in \eqref{m}, $T_0$ in \eqref{tsomega} and the definition of $b_i$, we have
\bea
\dot{T_0}&=&\ft{1}{2\pi}\Big(\fft{1}{(\epsilon +1)r_0^2}-\sum_{i}^{N}\fft{1}{r_0^2-a_i^2}\Big)\,,\qquad
\dot{m}=-4\pi r_0 m\,\dot{T_0}\,,\nn\\
\dot{b_i}&=&4b_i \sum_{j\neq i}^{N}\frac{r_0^3}{(r_0^2-a_i^2)(r_0^2-a_j^2)}\,.
\eea
We find that the expressions of $\dot f_i$'s are better expressed in cases of odd and even dimensions:
 \be
 \dot{f_i}=\left\{
 \begin{aligned}
 	&-\ft{2r_0(r_0^4+2r_0^2a_i^2-7a_i^4)}{(r_0^2-a_i^2)(r_0^2+a_i^2)^3}\,,\quad &D=\hbox{odd};\\
 	&-\ft{12r_0^2-4a_i^2}{(r_0^2-a_i^2)(r_0^2+a_i^2)^2}\,,\quad &D=\hbox{even}.
 \end{aligned}
 \right.
 \ee
We can now obtain the expression of the entropy correction  by substituting the relevant $\dot{T_0},\dot{m},\dot{b_i} $ and $\dot{f_i}$ into \eqref{dotG} and finally \eqref{cSJ}.

Similarly, for calculating $\Delta J_0^1$, we need to adopt the prime notation, which is a derivative with respect to $r_0$, with $dT_0=0$ and $d\Omega_0^{i\neq 1}=0$. This implies
\be
a_1^\prime=\frac{1}{2 r_0 a_1}\Big( a_i^2-r_0^2+(r_0^2+a_1^2)^2\big(\frac{1}{(\epsilon +1)r_0^2}-\sum_{j=2}^{N}\frac{1}{r_0^2-a_j^2}\big)\Big),
\qquad a_i^\prime=\dot{a_i},\quad i>1\,,
\ee
together with
\bea
\Omega_0^{1\prime}&=&\frac{(r_0^2-a_1^2)a_1^\prime}{(r_0^2+a_1^2)^2}
-\frac{2a_1r_0}{(r_0^2+a_1^2)^2}\,,\qquad
m^\prime = -2 a_1 m \Omega_0^{1\prime}\,,\nn\\
b_i^\prime &=& 2b_i\sum_{j\neq i}^{N}\frac{a_i a_i^\prime-a_j a_j^\prime}{a_i^2-a_j^2}\,.
\eea
The expression of $f_i^\prime$ is relatively more complicated, given by
 \bea
f_1^\prime&=&\left\{
\begin{aligned}
	&-\frac{2\left((a_1^5-14a_1^3r_0^2+9a_1r_0^4)a_1^\prime
+r_0(9a_1^4-14a_1^2r_0^2+r_0^4)\right)}{(a_1^2+r_0^2)^4}\,,\qquad &D=\hbox{odd;}\\
	&-\frac{16a_1r_0(a_1^2-2r_0^2)a_1^\prime
-4(a_1^4-8a_1^2r_0^2+3r_0^4)}{(a_1^2+r_0^2)^4}\,,\qquad &D=\hbox{even;}
\end{aligned}
\right.\\
f_i^\prime&=&\dot{f_i},\qquad i>1\,.
\eea
Similarly, we can incorporate the above equations into \eqref{primeG} and \eqref{cSJ} and then obtain all $\Delta J^i$.

\end{document}